# Title: Moiré-localized interlayer exciton wavefunctions captured by imaging its electron and hole constituents

**Authors:** Ouri Karni[1]†, Elyse Barré[2,3]†, Vivek Pareek[4]†, Johnathan D. Georgaras[5]†, Michael K. L. Man[4]†, Chakradhar Sahoo[4]†, David R. Bacon[4], Xing Zhu[4], Henrique B. Ribeiro[1,2], Aidan L. O'Beirne[6], Jenny Hu[1], Abdullah Al-Mahboob[4], Mohamed M. M. Abdelrasoul[4], Nicholas S. Chan[4], Arka Karmakar[4], Andrew J. Winchester[4], Bumho Kim[7], Kenji Watanabe[8], Takashi Taniguchi[9], Katayun Barmak[10], Julien Madéo[4], Felipe H. da Jornada[5], Tony F. Heinz[1,2], and Keshav M. Dani[4]∗

**Affiliations:**

[1] Department of Applied Physics, Stanford University; Stanford, California, 94305, USA.

[2] SLAC National Accelerator Laboratory; Menlo Park, California, 94025, USA.

[3] Department of Electrical Engineering, Stanford University; Stanford, California, 94305, USA.

[4] Femtosecond Spectroscopy Unit, Okinawa Institute of Science and Technology Graduate University, Onna-son; Okinawa, Japan 904-0495.

[5] Department of Material Science and Engineering, Stanford University; Stanford, California, 94305, USA.

[6] Department of Physics, Stanford University; Stanford, California, 94305, USA.

[7] Department of Mechanical Engineering, Columbia University; New York, New York 10027, USA.

[8] Research Center for Functional Materials, National Institute for Materials Science, 1-1 Namiki, Tsukuba 305-0044, Japan.

[9] International Center for Materials Nanoarchitectonics, National Institute for Materials Science, 1-1 Namiki, Tsukuba 305-0044, Japan.

[10] Department of Applied Physics and Applied Mathematics, Columbia University; New York, New York 10027, USA.

† These authors contributed equally to this work.

*Corresponding author. Email: kmdani@oist.jp

**Introductory paragraph:**

Interlayer excitons (ILXs) – electron-hole pairs bound across two atomically thin layered semiconductors – have emerged as attractive platforms to study exciton condensation,[1–4] single-photon emission and other quantum-information applications[5–7]. Yet, despite extensive optical spectroscopic investigations[8–12], critical information about their size, valley configuration and the influence of the moiré potential remains unknown. Here, we captured images of the time- and momentum-resolved distribution of both the electron and the hole that bind to form the ILX in a $WSe_2/MoS_2$ heterostructure. We thereby obtain a direct measurement of the interlayer exciton diameter of ~5.4 nm, comparable to the moiré unit-cell length of 6.1 nm. Surprisingly, this large ILX is well localized within the moiré cell to a region of only 1.8 nm – smaller than the size of the exciton itself. This high degree of localization of the interlayer exciton is backed by Bethe-Salpeter equation calculations and demonstrates that the ILX can be localized within small moiré unit cells. Unlike large moiré cells, these are uniform over large regions, thus allowing the formation of extended arrays of localized excitations for quantum technology.

**Main Text:**

Heterobilayers made by stacking different monolayers (1L) of transition-metal dichalcogenide (TMDC) semiconductors support ILXs, states in which the electrons and holes residing in separate layers bind to one another to form a neutral excitation [8]. The utilization of these long-lived and tunable excitations for quantum applications or for studying new many-exciton phases and interactions requires critical knowledge about the real- and momentum-space structure of the ILX: the valley character of its constituent electron and hole; its size, namely the extent of the electron distribution around its hole counterpart; and the degree of the ILX confinement within the moiré potential. These attributes are key to determining the nature of the light-matter interactions with the ILX, such as their direct bandgap character and their polarization selection rules, as well as the ability of the ILX to form uniform periodic arrays of localized quantum emitters in the moiré pattern [5]. Additionally, the ILX size and localizability determine their density-dependent many-body physics – at low densities strong center-of-mass (COM) confinement inhibits the formation of condensates [13], while at higher densities, the size and localization determine Mott dissociation thresholds[2] or the formation of multi-ILX complexes[3,4].

So far, studies on moiré-localized ILXs have been focused on large moiré unit cells due to the prevailing assumption that moiré confinement requires a moiré-period which is much larger than the ILX itself[9,14]. Such systems were studied by optical spectroscopy that could only access a narrow sector of the momentum-space distribution of the ILX wavefunction[9,10,12,15]. While such optical spectra were qualitatively consistent with a theoretical picture of ILXs spatially localized by the moiré-potential, these experiments did not yield the ILX size, and often relied on phenomenological moiré potentials that were not expected from first-principles calculations to deduce the presence of a moiré-localized ILX[9,16]. The valley configuration of the electron and hole of the ILX also remained contested[9,17]. Furthermore, such systems exhibit only small regions of uniform periodicity of the moiré lattice due to strain fields[18], and hence cannot support the formation of extended arrays of localized ILX that are the starting point of recent quantum technology proposals. Revealing those hidden properties together with a direct measurement of the extent of the ILX localization requires knowing the full momentum-space distributions of *both* of its constituent particles. Recent time- and angle-resolved photoemission spectroscopy experiments on microscopic samples (TR-µ-ARPES) measured the size of the exciton in 1L WSe$_2$ via the momentum distribution of only the constituent electron [19]. However, accessing the coordinates of the constituent hole has remained beyond experimental reach because of the challenge in measuring the small, momentum-resolved, photo-induced reductions in electron density over sufficiently large momentum-space in the 2D valence band (VB) to cover the hole wavefunction.

Here we image both the electron and hole momentum-space distributions of an ILX in a TMDC heterobilayer exhibiting a relatively small moiré pattern. From these distributions, we extract both the ILX size and its COM confinement, and find, surprisingly, that even though the ILX diameter is comparable to the moiré-period, its COM coordinate is tightly confined by the moiré-potential. Further, we observe the direct bandgap nature of the ILX with the electron and hole residing in the K-valleys of the two layers, and the anomalous negative dispersion of the photoemitted electron, establishing the excitonic origin of the signal. Finally, the hole images, which are acquired independently of the photoemission matrix elements, offer a direct quantitative measurement of the ILX density and exhibit a broadening in momentum distribution when their density surpasses one ILX per moiré site, akin to excitons in quantum dots. In all, these findings promote the prospect of using small-period moiré patterns, which are homogeneous and more robust against strain than the large moiré lattices[18,20], to host arrays of quantum emitters.

The studied heterobilayer consisted of 1L $WSe_2$ placed on top of 1L $MoS_2$ with a 2.2±0.8° twist-angle between their crystallographic axes. Together with the 3.8 % lattice mismatch between the two layers this is expected to yield a 6.1 nm period for the moiré pattern[5], which would make it more homogeneous compared to the larger moiré periods of more lattice-matched heterobilayers[18,20]. To isolate the ILX from the conducting substrate, while still preventing the sample from charging during the photoemission measurement, a thin hBN layer was placed on top of a conducting silicon substrate prior to stacking the sample [19,21] (see Fig. 1d, and Methods, for fabrication details). Confirming the coupling between the $WSe_2$ and $MoS_2$ layers, the photoluminescence (PL) spectrum from the sample (Fig. 1b) exhibits an ILX peak at 1 eV[11], and the optical reflection spectrum of the sample exhibits the expected signatures of the moiré pattern (Fig. S1).

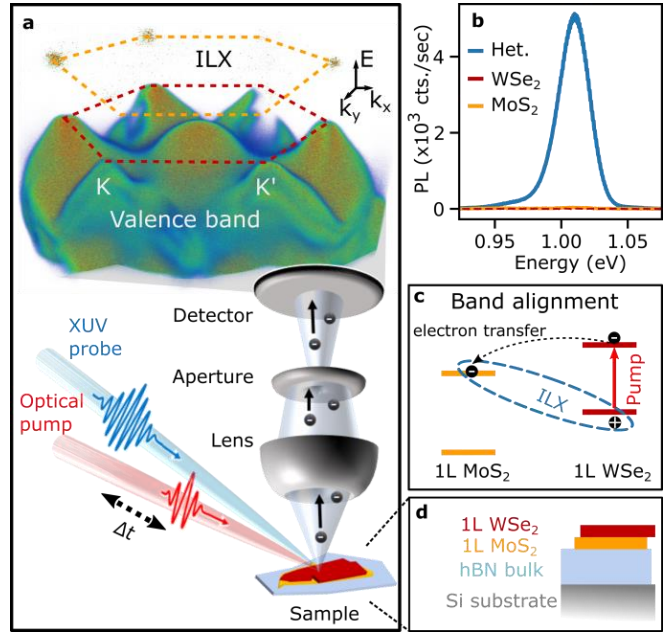

**Fig. 1 - Sample structure and experimental configuration. a.** Schematic representation of the TR-µ-ARPES setup and sample structure. The XUV probe pulse impinges on the sample at a delay Δt relative to the optical pump pulse. Photoemitted electrons from the excited sample are filtered through a micron-scale spatially selective aperture and resolved in energy and momentum by a 3D time-of-flight detector. We obtained a volumetric dataset of the band-structure of the entire Brillouin zone with an energy range covering the relevant VB and the photoexcited electron features (Top). **b.** Low temperature (10 K) infrared PL of the 1L $MoS_2$ (yellow), $WSe_2$ (red), and the heterobilayer (blue). **c.** Schematic of the band-edge alignment of the layers and the electron-transfer process leading to formation of the ILX. **d**. The layer arrangement of the sample heterostructure.

We first characterized the unperturbed band-structure of the heterobilayer (prior to excitation) at 100 K using our previously reported TR-µ-ARPES capabilities [19,21], based on a photoelectron momentum microscope [22,23] in combination with a table-top source of femtosecond XUV pulses at 21.7 eV (see Fig. 1a, and Methods for details). We observe a band-structure consistent with previous calculations (Fig. 2a) [11,24]. The VB maxima (VBM) appear at the K valleys (defining the zero of our energy scale), with both the upper and the lower spin-orbit split bands of $WSe_2$ clearly visible and separated by 0.5 eV. At lower energies (-0.85 eV) a single broad band is seen (Fig S2), corresponding to the two spin-orbit split VBs of $MoS_2$ that are unresolvable in our measurement due to inhomogeneous broadening. At the Γ valley, we observe two hybridized bands. Figure 2b presents a view of the band-structure in 2D momentum space integrated over a small energy range (-0.2 to 0.2 eV) around the VBM. We clearly see the photoemission from the Γ, the K, and the K' valleys. The variations in photoemission intensity across the image originate from the differing photoemission matrix elements [25], which depend sensitively on the incidence angle and the polarization [26] of the XUV pulse.

Having identified the signatures of the heterobilayer in equilibrium, we turned to study the ILX and the momentum-resolved distributions of its electron and hole by photoexciting the sample resonantly at the A-exciton in $WSe_2$ with a 170 fs, 1.67 eV pump pulse. A rapid transfer of the electron from $WSe_2$ to $MoS_2$ results in the formation of the ILX (schematic in Fig. 1c). After a time-delay (Δt) of a few tens of picoseconds, we applied the XUV probe pulse to record the momentum- and energy-resolved photoemission spectra from the photoexcited heterobilayer and follow its evolution as the ILX cooled and reached quasi-equilibrium.

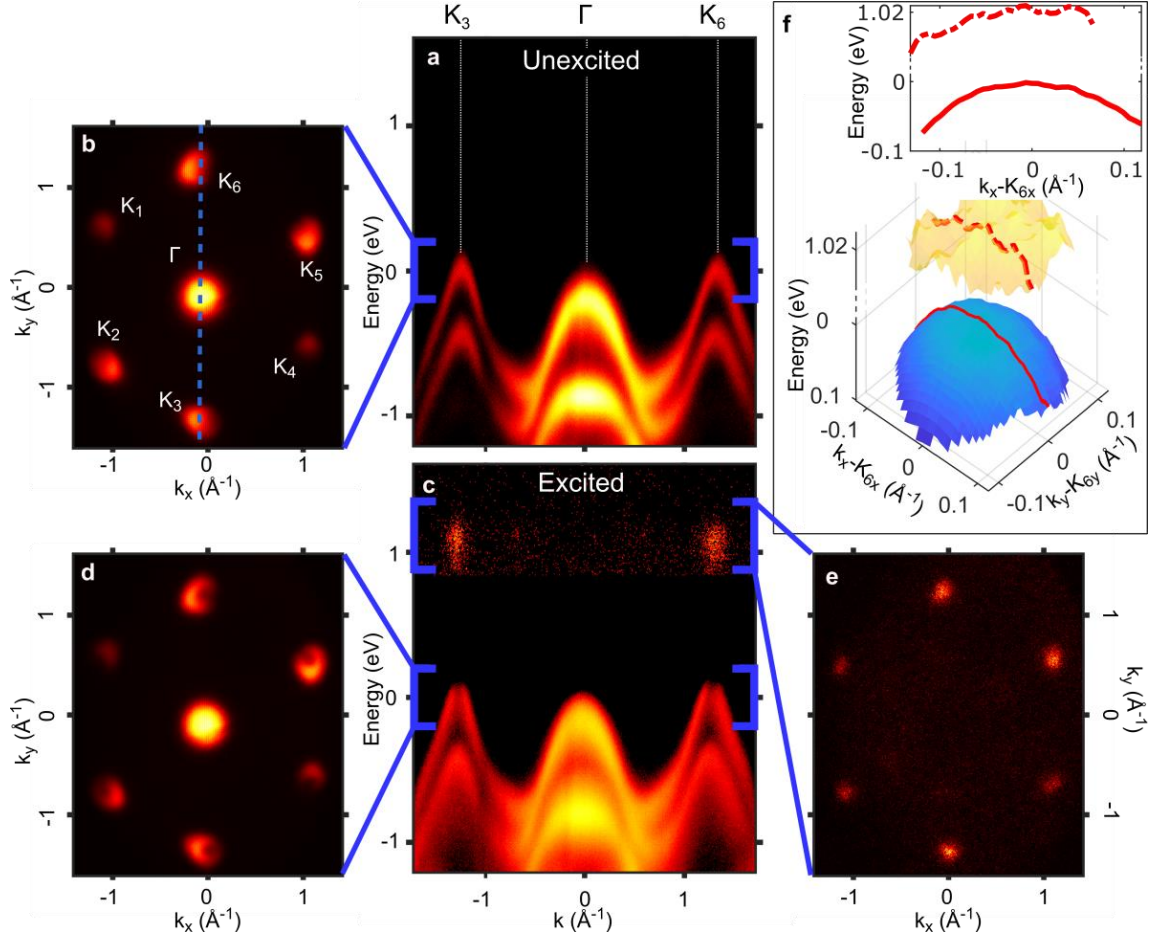

**Fig. 2 - Static and photoexcited TR-µ-ARPES measurements. a, c.** Energy-momentum cuts along the K-Γ direction without (a) and 25 ps after (c) optical excitation. **b, d, e.** Momentum-space images around VBM (b), as well as 25 ps after photoexcitation around VBM (d) and around the ILX energy (e). These ARPES intensity maps are formed by integrating over the energy ranges defined by the blue brackets in (a) and (c). Clear signatures of the constituent electron and hole at 1 eV and at the VBM, respectively, can be seen after photoexcitation (c, d and e) in all K valleys. In (b), the dashed line marks the $K_3$-Γ-$K_6$ cut used in (a) and (c). In (c), the counts above 0.85 eV are enhanced by a factor of 50 for visibility. **f.** Emission from the constituent electron of the ILX exhibiting the anomalous negative dispersion [19] of the VB. This establishes the bound nature of the observed electrons and holes. Inset: Line-cuts of the dispersion of the constituent electron (dashed) and VB (continuous) along $k_x$ axis.

The signatures of the ILX are immediately recognized (Fig. 2c, e) in the TR- µ-ARPES data: Photoemitted electrons appear in all the corners of the Brillouin zone (BZ) 1eV above the VBM. This energy agrees well with the PL associated with the ILX. Furthermore, the dispersion curve of the photoemitted signal at 1 eV clearly shows a negative curvature similar to the negative dispersion of the K (and K') valley VB of $WSe_2$ (Fig. 2f). This anomalous dispersion, recently demonstrated for intravalley excitons in 1L $WSe_2$, is a hallmark of the excitonic origin of the ARPES response [19,27]. We note that this signal is qualitatively different from measuring the replica of the VB within a Bloch-Floquet picture [28], which requires a coherent pump that overlaps with the probe. Besides this signal, we observe a striking depletion of the ARPES signal around the $WSe_2$ VBM. This depletion, seen in all K and K' valleys (Fig. 2d), corresponds to the holes created in $WSe_2$ as part of the ILX. This momentum-resolved distribution of the hole within an exciton provides a new route to understand excitons, well beyond recent experiments that measure only the properties of the electron [19].

To obtain the quantitative 2D momentum distributions of the holes, we compared TR-µ-ARPES spectra taken before and after excitation (Fig. S3), which allowed us to obtain the hole momentum-space images independent of the k-dependence of the photoemission matrix elements (See Supplementary Material for details). At long delays, when excitons are in quasi-equilibrium at the lattice temperature, the depleted signal at wavevector **k** in the VB provides the probability of finding a hole with momentum $-\hbar\mathbf{k}$ (See Supplementary Material for details). We note that integrating this quantity over all k-space provides a direct measurement of the ILX density at any given delay (Fig. 3d) – an important quantity that generally can only be estimated in time-resolved optical experiments. In addition to momentum-space images of the hole distribution, we also obtained the momentum-resolved images of the electron distribution directly from the photoexcited TR-µ-ARPES spectrum [19]. We extracted quantitative parameters at different time delays by fitting these momentum-space electron and hole images to Gaussian distributions (Fig. 3a).

One of the extracted parameters is the Gaussian widths of these distributions in momentum-space, which provide important information about the ILX. This can be understood by noting the wavefunction of an idealized ILX, unconfined by any moiré potential, and with COM momentum $\hbar \mathbf{Q}$, is given by $|X^{\mathbf{Q}}\rangle = \sum_{vc\mathbf{k}} A_{vc\mathbf{k}}^{\mathbf{Q}} c_{c,\mathbf{k}+\alpha\mathbf{Q}}^{\dagger} c_{v,\mathbf{k}-\beta\mathbf{Q}} |0\rangle$, where $\alpha = \frac{m_e}{m_e+m_h}$ and $\beta = \frac{m_h}{m_e+m_h}$. Here $A_{vc\mathbf{k}}^{\mathbf{Q}}$ is the envelope function of the unconfined ILX, $c_{n,\mathbf{k}}^{\dagger}$ is the creation operator of an electron in band $n$ with wavevector $\mathbf{k}$, and $|0\rangle$ is the un-photoexcited ground state of the system[27]. For $\mathbf{Q} = 0$, the momentum distributions of the electron and hole are identical and directly provide the size of the exciton, *i.e.*, the relative electron-hole coordinate in real space (Fig. 3b) [19,27,29]. For finite $\mathbf{Q}$, the electron and hole distributions are mainly offset in momentum by different amounts, $\alpha\mathbf{Q}$ and $\beta\mathbf{Q}$, respectively. Thus, a distribution of excitons with finite $\mathbf{Q}$ yields electrons and holes displaying distinct momentum distributions (Fig. 3c). In general, there are two processes that may lead to a distribution of excitons with finite $\mathbf{Q}$: thermal effects and spatial confinement. For sufficiently low temperatures – smaller than the energy gap between different moiré-exciton bands, but still larger than the moiré-exciton bandwidth – thermal effects give rise to a small and temperature-independent broadening contribution that can be neglected (see Supplemental Materials). This is indeed the case in our experiments, as the extents of the electron and hole distributions display no meaningful temperature dependence (see Fig. S4). In contrast, the moiré potential localizes the COM coordinate of the ILX, resulting in a distribution of finite values of **Q**, which is encoded in the inequivalent electron and hole momentum distribution. We note that the above discussion assumes low-enough exciton densities to avoid multi-exciton and state-filling effects, as well as sufficiently long time-delays to allow the excitons to cool to the lattice temperatures. In our measurements, these conditions are achieved at photoexcitation densities below $3 \times 10^{12}$ cm$^{-2}$ (Fig. 3d,e), corresponding to less than one ILX per moiré cell, and after a few tens of picoseconds (Fig. 3f), at which point the hole distribution narrows to a width of $\sigma_k^h = 0.042 \pm 0.002$ Å$^{-1}$, while the electron momentum distribution exhibits a width of $\sigma_k^e = 0.061 \pm 0.003$ Å$^{-1}$.

To deduce the ILX size and spatial confinement from the momentum distributions of the electron and hole measured by TR-μ-ARPES, we express the moiré-localized ILX $|X_m\rangle$ as a linear combination of unconfined interlayer excitons $|X^\mathbf{Q}\rangle$ with different COM momenta $\hbar\mathbf{Q}=\hbar\mathbf{G}$, $|X_m\rangle = \sum_\mathbf{G} C(\mathbf{G})|X^\mathbf{G}\rangle$, where **G** is a reciprocal-lattice vector of the moiré unit cell and $C(\mathbf{G})$ are expansion coefficients. The measured intensity distributions of the electron and hole that constitute the ILX, $I_e(\mathbf{k})$ and $I_h(\mathbf{k})$, are expressed in terms of the expansion coefficients $A_{vc\mathbf{k}}^\mathbf{Q}$ and $C(\mathbf{G})$, as $I_e(\mathbf{k}) = \sum_\mathbf{G}\left|C(\mathbf{G})A_{\mathbf{k}-\alpha\mathbf{G}}^\mathbf{G}\right|^2$ and $I_h(\mathbf{k}) = \sum_\mathbf{G}\left|C(\mathbf{G})A_{\mathbf{k}+\beta\mathbf{G}}^\mathbf{G}\right|^2$. For Wannier-like excitons, the simultaneous measurement of these distributions allowed us to extract both $A_{vc\mathbf{k}}^{\mathbf{Q}=0}$ and $C(\mathbf{G})$. Their Fourier transforms provided the envelope functions of the relative electron-hole separation $\mathbf{r}$ as $\psi(\mathbf{r})$, and the COM coordinate **R** as $C(\mathbf{R})$, respectively. With this, one can approximate the envelope of the two body exciton wavefunction as a product of $\psi(\mathbf{r})$ and $C(\mathbf{R})$, with the real-space exciton radius and its COM localization being inversely proportional to the extent in momentum-space of $|A_{vc\mathbf{k}}^{\mathbf{Q}=\mathbf{0}}|^2$ and $|C(\mathbf{G})|^2$, respectively (see Supplementary Materials for calculation details). Accordingly, we deduce an experimental ILX radius (*i.e.*, the root-mean-square (RMS) radius of its relative electron-hole separation distribution) of 2.7 nm (5.4 nm diameter), consistently larger than the 1.4 nm size of the intra-layer exciton in WSe$_2$[19]. We also obtain an RMS localization of the COM distribution of 0.9 nm, revealing an ILX whose COM coordinate is rather tightly pinned to the minimum of the moiré potential landscape.

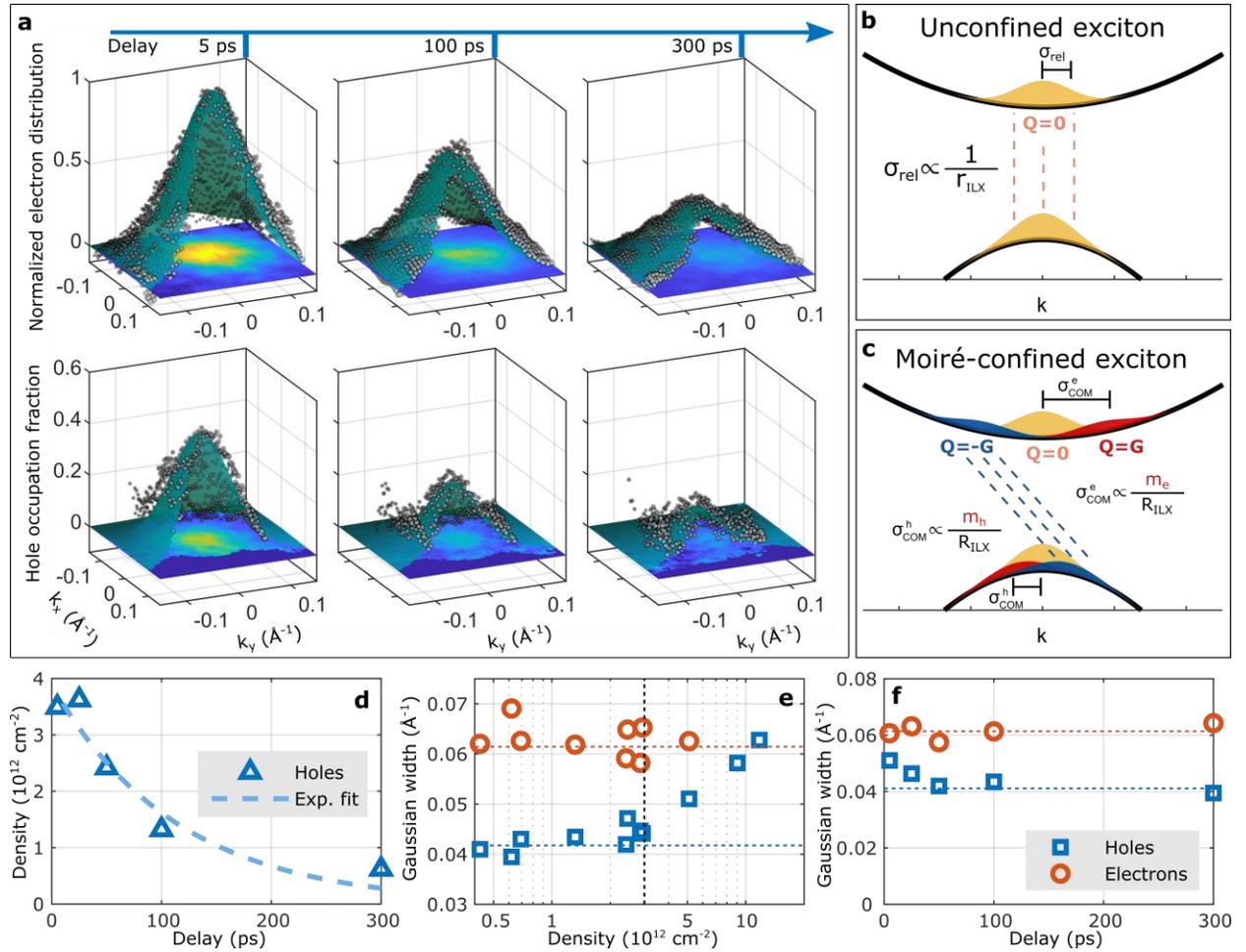

**Fig 3 – Time- and momentum-resolved distributions of the ILX electrons and holes. a.** Distributions around the K valley ($K_6$ from Fig. 2). Point clouds and heat maps represent the data. Teal surfaces are Gaussian fits. The coordinate $(k_x,k_y)=(0,0)$ is set to the VB maximum. Delay values are indicated above each panel. The normalized photoexcited electron distribution (top row) exhibits a much larger width than the hole distribution (bottom row). **b.** Schematic of the momentum distributions (yellow) electron and hole constituting an unconfined exciton with zero COM momentum, showing identical extents. The black curves indicate the conduction band and VB with their different curvatures. **c.** Similar schematic for a confined exciton, composed of multiple non-zero Q states (blue, yellow, and red). The electron and hole each carry only a fraction of Q that is proportional to their respective effective mass, resulting in unequal momentum distributions as shown. **d.** Hole density, indicative of ILX density, as function of delay. The exponential fit yields a lifetime of $113 \pm 34$ ps, much longer than that of intralayer excitons. **e.** The quasi-equilibrium photoexcited electron and hole distribution widths at different densities, indicating the onset of a broadening at $\sim 3 \times 10^{12}$ cm$^{-2}$, corresponding to more than one ILX per moiré lattice. **f.** Width of the momentum distribution of the electron and hole versus time-delay showing the two distinct values for quasi-equilibrium at long-time delays.

Our experimental results on the spatial size and localization of the ILX in a well-characterized sample motivate us to seek deeper physical insight into the nature of moiré-confined ILXs using state-of-the-art many-body perturbation theory calculations. We have accordingly carried out a set of first-principles and effective-Hamiltonian calculations. We first performed *ab initio* GW plus Bethe-Salpeter equation calculations [30,31] of the **Q** = 0 ILX wavefunction distribution in an artificially commensurate MoS$_2$/WSe$_2$ heterobilayer to obtain $A_{vc\mathbf{k}}^{\mathbf{Q}=\mathbf{0}}$. This yielded an RMS radius of 2.4 nm (4.8 nm diameter) in real space (See Fig. 4a). A further calculation employing an effective moiré ILX Hamiltonian[9,14] yielded a COM coordinate that is localized in real-space to within an RMS radius of 1.3 nm around the moiré potential minimum (See Fig. 4b). Our calculations show that both $|A_{vc\mathbf{k}}^{\mathbf{Q}=\mathbf{0}}|^2$ and $|C(\mathbf{G})|^2$ are well-described by Gaussian distributions, and both the ILX size and its COM localization are in reasonable agreement with our experiment (see Methods for details).

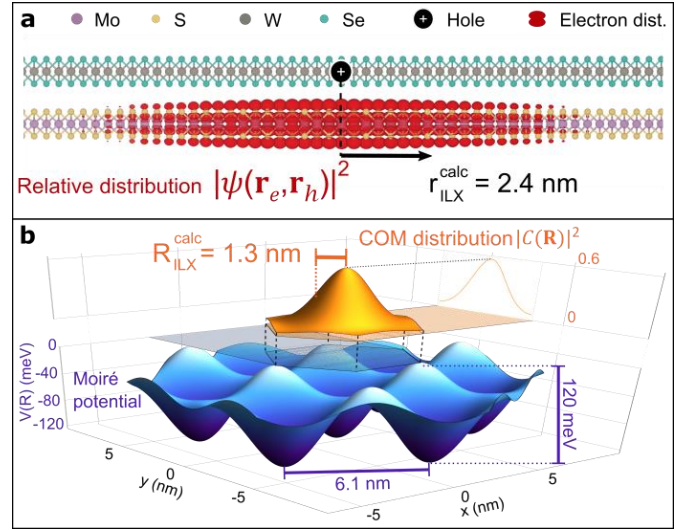

**Fig. 4 - ILX wavefunctions in real-space. a.** Modulus squared of the ILX wavefunction $|\psi(\mathbf{r}_e, \mathbf{r}_h)|^2$ (red) as a function of the electron coordinate $\mathbf{r}_e$, for the hole coordinate $\mathbf{r}_h$ fixed as indicated in black. The moiré confinement was not included in this plot. **b.** The calculated real-space envelope function squared for the COM coordinate of the ILX, $|C(\mathbf{R})|^2$, on top of the calculated spatial landscape of the moiré potential V($\mathbf{R}$).

Our simultaneous measurement of the electron and hole momentum distributions in a bound state provides previously inaccessible information about both the relative and COM coordinates that characterize the ILX two-body wavefunction. A measurement of the relative coordinate accesses the envelope function of the exciton and provides, for the first time, its size – a critical parameter in determining the nature of and thresholds for various many-exciton processes, such as the Mott transition [32] and exciton-exciton annihilation[33]. For example, the measured ILX size suggests a Mott transition density[32] in the range of $10^{12}$ to $10^{13}$ cm$^{-2}$, which coincides with a density of a few excitons per moiré unit cell. It thus raises interesting questions about the interplay of the moiré potential and the Mott transition. The COM coordinate of the exciton also plays an important role in understanding the underlying physics of excited states. The observed localization, taken together with the onset of momentum-space broadening as the ILX density crosses one per moiré cell, resembles the behavior of excitons in impurity-based quantum dots[3,4]. Such observations support emerging schemes in quantum technology that utilize an array of quantum-dot-like states created by the periodic moiré potential [5,7]. Compared to larger moiré periods, our use of lattice-mismatched heterobilayers with a small moiré period possesses the crucial benefit of extended spatial homogeneity and robustness against strain fields[18]. For further studies, the COM distribution also provides direct access to the temperature of an ensemble of excitons. Combined with the ability to read the exciton density directly from the data, such wavefunction measurements offer an important new tool to study of many-body excitonic states, their interactions with correlated electrons[34], and the corresponding phase diagrams.

**Methods**

Sample fabrication

1L MoS$_2$ (2D Semiconductors) and 1L WSe$_2$ (flux-grown, Columbia University, NY, USA) flakes were exfoliated onto transparent polydimethylsiloxane (PDMS) stamps and identified by their optical contrast and their photoluminescence (PL). Hexagonal boron nitride (hBN-National Institute for Material Science, Japan) was directly exfoliated on a bare silicon wafer (with a 1-2 nm thick native oxide). A flat and thin hBN flake of 18 nm thickness (as measured by AFM) was located by optical contrast. The heterostructure (clear area of 25x8 µm$^2$) was assembled by dry-transfer of the monolayer flakes onto the hBN, and then vacuum-annealed for 2 hr at 150° C. The crystallographic alignment of 1L WSe$_2$ and 1L MoS$_2$ was established judging by the sharp edges of the flakes. After fabrication, a polarization-resolved second-harmonic generation measurement determined the twist angle to be 2.2±0.8°.

Optical characterization of the samples

The PL spectra of the monolayers and their heterostructure were measured in a home-built confocal microscope setup. The sample was held at a controlled temperature in a cryostat (Montana Instruments). Laser light at 532 nm wavelength was used for excitation and was focused to a diffraction-limited spot (~1 µm) on the sample by a microscope objective (Olympus LCPLN50xIR). The infrared PL was collected by the same objective; it was detected and analyzed by a Jobin Yvon iHR550 spectrometer equipped with a Spectrum One InGaAs array. For reflection-contrast measurements (shown in the Supplementary Material) white light from a stabilized tungsten-halogen light source (Thorlabs SLS201L) was delivered to the sample, and its reflection collected, through the same optical setup. To analyze the reflection spectrum in the visible range a Synapse CCD camera was employed as a detector.

Time- and angle-resolved photoemission spectroscopy

Time- and angle-resolved photoemission spectroscopy is performed by coupling ultrafast pulses with an angle-resolved, energy-resolved electron microscope (Metis 1000, SPECS GmbH) [19,21]. To monitor carrier dynamics, a pump-probe scheme has been developed with a tunable optical pump pulse that photo-excites the sample and an XUV probe pulse that photo-emits the electrons. Both pulses originate in a Yb-doped fiber laser producing 250 fs, 50 µJ pulses at a wavelength of 1030 nm and a repetition rate of 2 MHz. The optical pump pulse, at 1.67 eV (740 nm), is produced by a non-colinear optical parametric amplifier (NOPA) pumped by part of the fundamental laser power. The high-energy XUV photons are produced by high-harmonic generation (HHG), achieved by frequency doubling the remaining of the fundamental laser light, and then focusing it into Kr gas. One of the high harmonics, at an energy of 21.67 eV is selected using a combination of Al and Sn filters. Both the pump and the probe are focused onto the region of interest on the sample. The pump energy was tuned in resonance with one of the absorption peaks of $WSe_2$ in the heterobilayer at the relevant experimental temperatures ranging between 80 K to 100 K. The probe energy is sufficient to photo-emit electrons with momenta covering the entire BZ of the sample, which emerge with different energies and at different angles. The photoelectrons are guided by the electrostatic lenses of the electron microscope and are detected by a time-of-flight detector, which measures the energy of the emitted electrons. Finally, the electrons are imaged either in real space (spatial- and energy-resolved) or in *k*-space (momentum- and energy-resolved) for different pump-probe time delays. More details of the TR- µ-ARPES instrumentation can be found in Ref. [19,21]

*Ab initio* calculations

We performed our mean-field density-functional theory (DFT) calculations with a Perdew-Burke-Ernzerhof exchange-correlation functional using the Quantum Espresso code [35]. The calculations were done on a unit cell in the plane-wave basis using the Optimized Norm-Conserving Vanderbilt pseudopotentials from the SG-15 dataset [36]. We first considered a unit cell for an artificially periodic $MoS_2$/$WSe_2$ heterobilayer, for which we take the lattice constant as the average of the individual $MoS_2$ and $WSe_2$ experimental lattice constants, 3.22 Å. The distance between repeated unit cells in the out-of-plane direction was taken as 25 Å and the Coulomb interaction was truncated [37] to effectively remove interaction between the repeated simulation cells along the out-of-plane direction.

We relaxed the atomic positions for the $MoS_2$/$WSe_2$ heterostructure in the $R_h^X$ stacking [5] including van der Waals interactions within the frameworks of Dione *et al.* [38] and Cooper [39], and obtained a separation of 6.2 Å between the center of $MoS_2$ and $WSe_2$ layers. The dielectric matrices used to solve the Bethe-Salpeter equation (BSE) to obtain excitonic states were evaluated up to a cutoff of 6 Ry on a 90x90x1 Monkhorst-Pack *k*-point grid and computed with the BerkeleyGW software package [40].

To solve for the Bethe-Salpeter equation (BSE), we first evaluated the electron-hole interaction kernel on a coarse 90x90x1 *k*-grid with a fully relativistic framework, which was then interpolated onto a patch of a 600x600x1 *k*-point grid around the K-valley. We diagonalized the BSE effective Hamiltonian with the BerkeleyGW package keeping four CBs and four VBs. Similarly to previous calculations[41], the sample patch only includes the 9593 *k*-points within 0.2 Å$^{-1}$ of the K point of the Brillouin zone and properly includes all the qualitative features of the desired interlayer excitons.

Model-Hamiltonian calculations of the exciton confinement

In the absence of any moiré effects, the excitons that we computed from first principles in the previous section can be labeled with respect to a principal band and wavevector (**Q**) quantum numbers. To a good approximation, because of vanishingly small exchange terms for ILX, these excitons can be described with an effective-mass approximation. The energy of the lowest moiré-unconfined ILX is given by $E_\mathbf{Q} = E_0 + \hbar^2 Q^2/(2M)$, where $E_0$ is the energy of the lowest ILX and $M$ is the total exciton mass. However, the presence of a moiré pattern with a periodicity larger than the lattice constants of the constituent monolayers allows for multiple excitons with distinct wavevectors (and, to lesser extent, band indices) to mix. This mixing can be rationalized in terms of an effective moiré potential, $\Delta(\mathbf{R})$, that describes the local variation of the bandgap across the moiré unit cell stemming from a different atomic registry at each point **R**.

We follow refs. [14] and [9] and describe the effective Hamiltonian for an ILX with wavevector **q** inside the moiré Brillouin zone that experiences a moiré potential as

$$H_\mathbf{q} = E_0 + \frac{\hbar^2|\mathbf{q} - \kappa|^2}{2M} + \Delta(\mathbf{R}),$$

where the momentum offset of $\kappa$ ($\kappa$ is the K point of the hexagonal moiré Brillouin zone) is due to the finite twist angle between the two individual TMDC layers [42].

The moiré potential energy is a smooth periodic function with extrema at specific high-symmetry points of the moiré lattice and can be approximated in the lowest-order harmonic expansion [9,14] as

$$\Delta(\mathbf{R}) \approx \sum_{j=1}^{6} V_j \exp(i\mathbf{G}_j \cdot \mathbf{R}),$$

where $\mathbf{G}_j$ are the first-shell reciprocal lattice vectors of the moiré Brillouin zone [9,14]. We first fit the potential parameter $V_j$ by performing DFT calculations on a series of high-symmetry stacking configurations and obtain a total potential width of 125 meV, similarly to previous calculations [5,43], and hence $|V_j|_{j=1..6} = 12.5\ meV$.

The model Hamiltonian expressed as such is solved in a plane-wave basis, where we retain moiré reciprocal-lattice vectors $\mathbf{G}$ with maximum magnitude of $|\mathbf{G}| = 10\mathbf{G}_1$. For each moiré-confined ILX with wavevector $\mathbf{q}$, we obtain the ILX envelope function $C_\mathbf{q}(\mathbf{G})$, as defined in the main text, by diagonalizing $H_\mathbf{q}$. The envelope function squared of the lowest-energy eigenstate can be approximated as an effective 2D Gaussian distribution whose width $\sigma_{COM}^r$ can be deduced from its momentum distribution – more specifically, from the variance $\langle|\mathbf{G}|^2\rangle_{\mathbf{q}=\kappa} \equiv \sum_\mathbf{G} |\mathbf{G}|^2 |C_{\mathbf{q}=\kappa}(\mathbf{G})|^2 = 2\sigma_{COM}^2$. We find numerically $\sigma_{COM} \approx 0.053\ Å^{-1}$. At 0 K, this width is directly related to the reciprocal of the COM localization radius (see Supplementary Material).

To account for thermal effects, we must consider the momentum distribution of $|C_\mathbf{q}(\mathbf{G})|^2$ for a range of values of $\mathbf{q}$ in the moiré Brillouin zone, the variance of which is related to an effective, thermally averaged width $\sigma_{COM,T}$. Because we are in a regime where $W < k_b T < \Delta E$, where $W$ is the bandwidth of the lowest-energy moiré exciton ($\sim$ 2 meV, from our calculations), $T$ is the temperature, and $\Delta E$ is the energy difference between the first and second moiré-exciton bands ($\sim$ 30 meV, from our calculations), we may assume that the lowest exciton band across the moiré Brillouin zone is uniformly occupied while higher bands are empty. From our numerical calculations, we find that finite-temperature effects increase $\sigma_{COM}$ to a value $\sigma_{COM,T}$ that is less than 10% larger. Furthermore, for sufficiently large $\sigma_{COM}$, one can approximate the thermal effects as $\sigma_{COM,T}^2 \approx \sigma_{COM}^2 + \sigma_T^2$, where $\sigma_T = \sqrt{7/6}\,|\mathbf{G}_1|/4 \approx 0.032$ Å$^{-1}$ is an effective thermal broadening contribution to the width of the COM distribution in reciprocal space and which only depends on the geometry of the moiré Brillouin zone.

**Supplementary Materials**

Materials and Methods

Supplementary Text

Figs. S1 to S6

**Acknowledgments:**

We thank the OIST engineering support section, and Yuki Yamauchi from OIST Facilities Operations and Use section for their support. Sample fabrication made use of the facilities in the Stanford Nano Shared Facilities (SNSF), supported by the National Science Foundation under award ECCS-1542152. The computational work was supported by the Center for Computational Study of Excited State Phenomena in Energy Materials, which is funded by the US Department of Energy (DOE), Office of Science, Basic Energy Sciences, Materials Sciences and Engineering Division, under Contract No. DE-AC02-05CH11231, as part of the Computational Materials Sciences Program. We acknowledge the use of computational resources at the National Energy Research Scientific Computing Center, a DOE Office of Science User Facility supported by the Office of Science of the US DOE under the above contract.

**Funding:**

AMOS program, Chemical Sciences, Geosciences, and Biosciences Division, Basic Energy Sciences, U.S. Department of Energy (TFH, OK, HBR)

Gordon and Betty Moore Foundation's EPiQS Initiative through grant number GBMF9462 (JH, ALO)

Koret Foundation (OK)

Natural Science and Engineering Research Council (NSERC) of Canada, fellowship PGSD3-502559-2017 (EB)

Femtosecond Spectroscopy Unit (VP, MKLM, CS, DRB, XZ, AAM, MMMA, NSC, AK, AJW, JM, KMD)

JSPS KAKENHI Grant No. JP17K04995 (MM)

Kicks Startup Fund (KMD NSC)

JSPS KAKENHI Grant No. 21H01020 (JM)

FAPESP for post-doctoral fellowships, grant number 2018/04926-9 and 2017/20100-00 (HBR).

National Science Foundation Materials Research Science and Engineering Center DMR-1420634 and DMR- 2011738 (KB, BK)

Elemental Strategy Initiative, conducted by the MEXT, Japan, Grant No. JPMXP0112101001 (KW, TT)

JSPS KAKENHI Grant No. 19H05790 and JP20H00354 (KW, TT)

**Author contributions:**

OK, EB and KMD conceived the project. KW and TT, BK and KB supplied raw materials for sample fabrication. EB and OK, supported by ALO, HBR, and JH, fabricated the samples. VP, MKLM, and CS collected the data with assistance from DRB, XZ, AAM, MMMA, AK, AJW and JM. VP, MKLM, CS, NSC and XZ performed preliminary analysis. EB and OK analyzed the data. FHJ and JDG performed theoretical calculations. JDG was supervised by FHJ. ALO, EB, HBR, JH, and OK were supervised by TFH. KMD supervised the project. All authors contributed to discussions and manuscript preparations.

**Competing interests:**

J.M., M.K.L.M., and K.M.D. are inventors on a patent application related to this work filed by the Okinawa Institute of Science and Technology School Corporation (US 2020/0333559 A1 published on October 22, 2020). The authors declare no other competing interests.

**Data and materials availability:**

All data are available in the main text or the supplementary materials.